\documentclass[pre,twocolumn]{revtex4}
\usepackage{graphicx}

\newcommand{\Db}{\mathbf{D}}

\newcommand{\Eb}{\mathbf{E}}

\newcommand{\rb}{\mathbf{r}}

\begin{document}

\title{Effects of electrostatic correlations on electrokinetic phenomena}
\date{\today}
\author{Brian D. Storey}
\address{Franklin W. Olin College of Engineering, Needham MA 02492}
\author{Martin Z. Bazant}
\address{Departments of Chemical Engineering and Mathematics, Massachusetts Institute of Technology, Cambridge, MA 02139}

\begin{abstract}
The classical theory of electrokinetic phenomena is based on the mean-field approximation, that the electric field acting on an individual ion is self-consistently determined by the local mean charge density. This paper considers situations, such as concentrated electrolytes, multivalent electrolytes, or solvent-free ionic liquids,  where the mean-field approximation breaks down.  A fourth-order modified Poisson equation is developed that captures the essential features in a simple continuum framework. The model is derived as a gradient approximation for non-local electrostatics of interacting effective charges, where the permittivity becomes a differential operator, scaled by a correlation length. The theory is able to capture subtle aspects of molecular simulations and allows for simple calculations of electrokinetic flows in correlated ionic fluids, for the first time.  Charge-density oscillations tend to reduce electro-osmotic flow and streaming current, and over-screening of surface charge can lead to flow reversal.  These effects also help to explain the suppression of induced-charge electrokinetic phenomena at high salt concentations. 
\end{abstract}

\maketitle
\section{Introduction}

The classical theory of the electric double layer and electrokinetic flow near a charged
surface is over a century old and remains in wide use today ~\cite{lyklema2003}.
The classical theory has been extremely powerful in a number of diverse fields such as
colloidal science,  biophysics, micro/nanofluidics and electrochemistry.
While the usefulness of the classical electrokinetic theory is not in question,
there is a long history of recognizing the limitations and offering
new formulations \cite{large_acis,vlachy1999}.

The equations are built on a set of assumptions which are
clearly violated in various instances.
The classical theory was developed for  a surface in chemical  equilibrium with a dilute  solution
of point ions with a double-layer
voltage on the  order of the thermal voltage, ${kT}/{e} = 25$ mV ~\cite{hunter_book,lyklema_book_vol2,russel_book}.
Stern recognized in 1924 that the assumption of point ions
leads to predicted concentrations that are  impossibly high at modest voltages.
Stern introduced the idea of a molecular layer of finite size to
reduce (but not eliminate) this un-physical  divergence by imposing a distance of closest approach of ions to the surface  \cite{stern1924}.
In many practical situations when the surface  is unknown or uncontrolled,
the macro-scale observable quantities such as capacitance or fluid slip velocity
are fit with  effective   Stern layer properties  to bring the classic model into agreement with experiment.

There has been recent interest in including finite ion size
 effects into the continuum electrokinetic model to go beyond the simple Stern layer approach \cite{large_acis}. It is apparently not well-known that Stern proposed such
 an approach as the final (un-derived) equation in his 1924 paper \cite{stern1924}.
 One driver for  interest in steric effects   are applications where electrokinetic
phenomena are exploited in devices with electrodes placed  in direct contact with the fluid \cite{ramos1999,ajdari2000,iceo2004b,squires2009review}.
These ``induced-charge electrokinetic phenomena"~\cite{cocis2010} have
 shifted attention to a regime where double-layer voltage reaches several Volts $\approx 100\, kT/e$, a regime where the point ion  theory is
certainly invalid.
To account for finite sized ions, a variety of ``modified Poisson-Boltzmann equations'' (MPB)
 have been proposed \cite{kilic2007a,large_acis}.
  The simplest possible MPB model is the one
proposed (and subsequently forgotten)
by Bikerman in 1942 \cite{bikerman1942}, which is a continuum approximation
of the entropy of ions on a lattice~\cite{grimley1947}.
Such  modifications to the continuum theory
can predict an otherwise unexplained high frequency flow reversal in AC electroosmotic pumps \cite{storey2008},
and capacitance of surfaces with no adsorption \cite{large_acis}.

Extensions of the classical
electrokinetic theory are also required for room-temperature ionic liquids (RTILs).
RTILs typically have  large organic cations and similar organic or smaller inorganic anions and  hold  promise as solvent-free electrolytes for super-capacitors, batteries, solar cells, and electro-actuators \cite{silvester2006,freyland2008,armand2009,ito2008,bai2008,buzzeo2004,ye2001,bhushan2008}.
For these applications,  data for the RTIL/metal interface has typically been  interpreted  through models based on the classical theory despite the fact that  this dense mixture of large ions bears little resemblance to a dilute solution of point-like ions.
 Recently, Kornyshev ~\cite{kornyshev2007} stressed the importance of finite-sized ions and developed a theory equivalent to Bikerman's, where the bulk volume fraction can be tuned to describe electrostriction of the double layer.

In spite of some success in applying a theory
which accounts for steric hinderance in electrolytes at high voltage and RTILs, these models
are unable to describe short-range Coulomb correlations~\cite{levin2002}.
In many important situations, classical
theory breaks down due to strong correlations between nearby ions.
In concentrated solutions, systems with multivalent ions (relevant for biology), RTILs, or molten salts,
electrostatic correlations which go beyond the mean electrostatic potential become dominant.
Correlations generally lead to {\it over-screening} of a charged surface, where the first layer provides more counter-charge than required; the next layer then sees a  smaller net charge of the opposite sign, which it overscreens with excess co-ions; and so-on.

Such overscreening is usually studied with molecular dynamics simulations, Monte-Carlo simulations (MC), Density Functional Theory (DFT), or integral equation methods based on the statistical mechanics of charged hard spheres.
While these simulations are based on more  realistic  assumptions than classic theory, the complexity prohibits analytical progress and the  computational cost and complexity  can be high. In many applications we are interested in charging dynamics, fluid flow,  or other macroscale behavior where a simple model is needed.
 To date, essentially all modeling of electrokinetic flow has been based
on the mean field approximation, where the electric field acting on the ions
is self consistently determined by the mean charge density.

In this paper  we maintain a continuum formulation and develop a modified Poisson equation
which accounts for electrostatic correlation effects in diffuse electric double layers. This model is applicable to
concentrated or multivalent electrolytes, room temperature ionic liquids, and molten
salts.
Recently, we (along with A.A. Kornyshev) derived and applied this continuum model for
RTILs \cite{bazant2011}.
In that work, we found good agreement in terms of the double layer structure and the capacitance when compared
to molecular dynamics simulations.
In the present work, we present the derivation in detail and apply the same continuum model to electrolytes, where correlations become important at high salt concentration and with multi-valent ions. We also couple the modified electrostatic theory to the Navier-Stokes equations, as we (along with M. S. Kilic and A. Ajdari) recently proposed~\cite{large_acis}. From this theoretical framework, we compute electrokinetic flows beyond the mean-field approximation for the first time. The model predictions are also compared to molecular simulations and some experimental data. 

Before we begin, we emphasize that any attempt to develop and modify continuum models for 
molecular scale  phenomena is fundamentally limited.
Nevertheless, our goal  is to develop and test models  that are simple enough
to facilitate a better understanding of electrokinetics in macroscale experiments and devices.
In particular, we describe flows in correlated electrolytes and ionic liquids with only one new parameter,  an electrostatic correlation length.

\section{Continuum electrokinetic equations }

\subsection{ Classical mean-field theory }

The classic theory of electrokinetics
assumes a dilute solution of point ions.
The electrochemical potential, $\mu_i$, of the $i^{th}$ ionic species in an ideal dilute solution is,
\begin{equation}
\mu_i^{\mathrm{ideal}} = k T \mathrm{log}c_i + z_i e \phi
\label{eq:mu_ideal}
\end{equation}
where $k$ is Boltzmann's constant, $T$ is the temperature, $c_i$ is the concentration,
$z_i$ is the charge number, $e$
is the elementary charge and $\phi$ is the electric potential.
We relate the flux of each species, $\mathbf{F}_i$,  to the
gradient in the chemical potential and conservation of mass yields,
\begin{equation}
\frac{\partial c_i}{\partial t} =
           -\nabla \cdot \mathbf{F}_i =
           -\nabla \cdot \left( c_i \mathbf{u} - \frac{D_i}{ k T} c_i  \nabla {\mu_i},\right).
\label{eq:NP}
\end{equation}
where $D_i$ is the diffusivity and $\mathbf{u}$ is the mass averaged velocity.
It is important to remember that directly relating the flux of each species to its own gradient in
chemical potential is an assumption that is strictly only valid in dilute solutions. This relationship assumes that the
diffusivity tensor is diagonal.
The system is traditionally closed by making the mean field approximation in which the
electric potential satisfies the Poisson equation,
\begin{equation}
\label{eq:poiseq}
-\nabla \cdot \varepsilon \nabla \phi =\rho= \sum_i z_i e c_i ,
\end{equation}
where $\rho$ is the charge density and $\varepsilon$ is the permittivity.
Equations \ref{eq:NP}-\ref{eq:poiseq} are  typically referred to as the Poisson Nernst Planck (PNP) equations.
The PNP equations are coupled to the Navier Stokes (NS) equations for fluid flow, where an electrostatic force
density, $\rho \nabla \phi$, is added,
\begin{eqnarray}
\rho_m \left( \frac{\partial \mathbf{u}}{\partial t} + \mathbf{u}\cdot \nabla \mathbf{u} \right)&= &- \nabla p + \eta \nabla^2 \mathbf{u} - \rho  \nabla \phi, \\
\nabla \cdot \mathbf{u} & =& 0,
\label{eq:NS}
\end{eqnarray}
where $\eta$ is the viscosity, $\rho_m$ is the mass density, and $P$ is the pressure.
In the classical theory the fluid properties such as the viscosity and permittivity are usually taken as constants.

Solutions to equations  \ref{eq:NP} - \ref{eq:NS}
require boundary conditions. Boundary conditions  can vary depending on the physical situation.
Typically, the no-slip condition for fluid velocity is assumed, but  modifications can allow for slip at a solid surface. A common boundary condition  for the ion conservation equation is that  there is no flux of ions at a solid surface. However, in cases with electrochemical reactions or ion adsorption, other boundary conditions are required.

The boundary condition for the potential depends upon the physics of the interface. Our interest is
on metal electrode surfaces where one can  simply fix
the applied potential $\phi = \phi_0$ or allow for a thin dielectric layer (or compact Stern layer) on
the electrode surface through the mixed boundary condition~\cite{bazant2005},
\begin{equation}
\Delta\phi_S = \phi - \phi_0 = \lambda_S \hat{n}\cdot\nabla \phi - \frac{q_S}{C_S}, \label{eq:sbc}
\end{equation}
where $\lambda_S=\varepsilon h_S / \varepsilon_S$ is an effective
thickness of the layer, equal to the true thickness $h_S$ multiplied by the ratio
of permittivities of the solution $\varepsilon$ and the layer
$\varepsilon_S$, and $C_S=\varepsilon_S/h_S$ is its capacitance.  When applying (\ref{eq:sbc}) to a metal electrode, one can set $q_S=0$ to model the Stern layer as a thin dielectric coating of solvent molecules~\cite{macdonald1962}, while specific adsorption of ions would lead to $q_S\neq 0$.

While the PNP+NS formulation is  widely studied and widely used,  the mathematical solution
can be complicated.
In many cases we can make {\em mathematical} simplifications that allow for analytical progress or simple models to be derived from the PNP+NS starting point.
In this work, we are developing a {\em physical} modification to the  equations.

\subsection{ Modifications for chemical effects }

In a recent review article we (along with M.S. Kilic and A. Ajdari) discuss in detail
a number of ways in which the classical mean-field theory of electrokinetics breaks down
 and propose some simple modifications for large voltages and concentrated solutions \cite{large_acis}.
We stress that attempts to go beyond the classic equations have a long history  and
 refer the interested reader to Ref. \cite{large_acis} for a more complete account of the literature.

To account for various thermodynamic non-idealities in concentrated solutions, we can
extend the chemical potential by adding an excess term to that of the ideal solution,
\[
\mu_i = k T \mathrm{log} c_i + z_i e  \phi +\mu_{ex}.
\]
In the case of volume constraints for finite-sized ions, following Bikerman \cite{bikerman1942}, this excess chemical potential
could be written as
\begin{equation}
\mu_i^{ex} =  - kT \mathrm{ln}(1-\Phi)
\end{equation}
where $\Phi$ is the local volume fraction of ions. The same model of the excess chemical
potential can also be derived from the configurational entropy of ions in a lattice gas in the continuum limit, as first noted by Grimley and Mott~\cite{grimley1947}..
We attribute this model to J.J. Bikerman though it
 has been independently rediscovered at least seven times since then and was possibly first discussed by Stern in
 1924. Other approaches can be used to modify the chemical potential for volume constraints, such as Carnahan Starling equation of state for the entropy of hard-spheres in the local density approximation~\cite{dicaprio2003,biesheuvel2007,carnahan1969}.
Regardless of the model for steric volume constraints, these modifications all
allow the formation of a condensed layer of ions very close to the surface
at high voltage. This layer forms at high voltage as the classic theory allows for an impossibly high
density of ions.

Another modification we have discussed in detail is charge induced thickening, where
one supposes that the viscosity of the fluid depends upon the local charge density and typically increases in the inner part of the double layer (effectively moving the ``shear plane" of no slip away from the surface).
Charge induced thickening
provides a possible explanation for the decay in induced-charge electro-osmotic flow~\cite{cocis2010} that is observed in many experiments at high salt concentration and/or high voltage
\cite{large_new,large_acis}.   Below, we will argue that electrostatic correlations may also play a significant role in explaining the data. 

The permittivity
$\varepsilon$ of a polar solvent like water is usually taken as a constant  in (\ref{eq:poiseq}), but numerous models exist for field-dependent permittivity $\varepsilon(|\nabla\phi|)$, as discussed in ~\cite{large_acis}. The classical effect of dielectric saturation reduces the permittivity at large fields due to the alignment of solvent dipoles~\cite{bockris_book,grahame1950,macdonald1962,macdonald1987}, although an increase in dipole density near a surface may have the opposite effect~\cite{abrashkin2007}. 
A recent model which included excess ion polarizability demonstrated excellent agreement with experimental capacitance data on surfaces with no adsorption \cite{Hatlo2012}.

While these and many other modifications have been explored, in this work we only consider the additional chemical effect of finite ion size, so we can focus on novel effects of electrostatic correlations.

\subsection{ Simple modification for Coulomb correlations }

The most fundamental modification of the classical theory, which has resisted a simple treatment, would be to relax the mean-field approximation.
While the study of electrostatic correlations in electrolytes has a long history, we are not aware of any attempts to go beyond the mean-field approximation (\ref{eq:poiseq}) in dynamical problems of ion transport or electrokinetics. Dynamical problems with bulk flow  would seem to require a simple continuum treatment of correlation effects, ideally leading to a general modification of Eq. \ref{eq:poiseq}.

In recent work on RTIL, we (along with A.A. Kornyshev) derived a Landau-Ginzburg type 
continuum model which accounts for
electrostatic correlations in a very simple and intuitive way \cite{bazant2011}.  
A general derivation based on nonlocal electrostatics will be developed in the next section, but first
we present the final result, which is a modified fourth-order Poisson equation,
\begin{equation}
 \nabla \cdot \Db  \equiv \varepsilon (\ell_c^2 \nabla^4 \phi -  \nabla^2 \phi) = \rho   \label{eq:Poisson}
\end{equation}
and a modified electrostatic boundary condition,
\begin{equation}
 \hat{n}\cdot\Db \equiv \hat{n}\cdot\varepsilon(\ell_c^2 \nabla^2 - 1)\nabla \phi=q_s
  \label{eq:PBC}
\end{equation}
where $\Db$ is the displacement field.
Due to Coulomb correlations, the effective permittivity $\hat{\varepsilon}$, defined by $\Db = - \hat{\varepsilon}\nabla\phi$, is a linear differential operator,
\begin{equation}
\hat{\varepsilon} = \varepsilon \left( 1 - \ell_c^2 \nabla^2 \right).  \label{eq:eps}
\end{equation}
This unusual dielectric response, signifying strong correlations, is consistent with some well known properties for molten salts, although we extend it here to more general situations.  In particular, for small, sinusoidal perturbations of the electric field of wavenumber $k$, the corresponding small-$k$ expansion of the Fourier transform of the permittivity,
\begin{equation}
\hat{\varepsilon}(k) = \varepsilon \left[ 1 + (\ell_c k)^2 \right]
\end{equation}
grows with $k$  in the case $\alpha_0>0$ where correlations promote charge density oscillations and discrete cation-anion-cation-anion... ordering. This matches known results for molten salts, although at smaller wavelengths (larger $k$) the permittivity transform $\hat{\varepsilon}(k)$ has divergences due to electronic relaxation and other phenomena~\cite{tosi1991,rovere1986}.  Here, we do not use the notion of wavelength-dependent permittivity, which only applies to small periodic bulk perturbations.  Instead, we introduce the concept of a {\it permittivity operator} in Poisson's equation, which can be applied to general nonlinear response in asymmetric geometries and near surfaces.
The new parameter $\ell_c$ is an effective length scale over which correlation effects are
important, discussed below. Its value is not precisely known, though we can place approximate bounds on its value.

Similar higher-order Poisson equations have been derived as approximations for the equilibrium statistical mechanics of point-like counterions (one component plasma) near a charged wall~\cite{santangelo2006,weeks2006,hatlo2010}; Santangelo~\cite{santangelo2006} showed that (\ref{eq:Poisson}) is exact for both weak and strong  coupling and  a good approximation at intermediate coupling with $\ell_c$ set to the Bjerrum length; Hatlo and Lue~\cite{hatlo2010} developed an approximation for $\ell_c$.  The extension to electrolytes and non-ideal solutions was first proposed in our review paper~\cite{large_acis} as part of a general modeling framework for electrokinetics, but without a derivation or any example calculations.   In our recent work on RTILs~\cite{bazant2011}, we presented a general variational derivation of the model and first applied this modified Poisson equation to predict double layer structure and capacitance (RTIL), using the ion size as the correlation
length scale.

Since Poisson's equation (\ref{eq:Poisson}) is now  fourth-order, we need an additional boundary condition. For consistency with our derivation below,  we neglect correlations very close to the surface (at the molecular scale) and apply the standard boundary condition, $-\varepsilon\hat{n}\cdot\nabla\phi = q_s$.  Equation (\ref{eq:PBC}) then implies
\begin{equation}
\hat{n}\cdot \nabla (\nabla^2 \phi) = 0  \label{eq:extraBC}
\end{equation}
which requires that the mean-field charge density is ``flat" at the surface.
Although this boundary condition is consistent with the  derivation of our model, we stress that it is neither unique nor rigorously derived. Alternative boundary conditions should be considered in the future, including the possibility of nonlocal models (e.g. which are required to describe density oscillations resulting from packing constraints~\cite{levin2002}).  Here, we use Eq.~(\ref{eq:extraBC}) partly for its elegant simplicity and partly since it seems to consistently produce remarkably good results with our model in comparison to molecular simulations, not only for RTIL~\cite{bazant2011}, but also for concentrated electrolytes, as described below.

\section{Derivation of the Modified Poisson Equation }

The following derivation is adapted from the supporting information of our recent publication with A. A. Kornyshev~\cite{bazant2011}, providing some more details and explanations of the steps.

\subsection{ Electrostatic energy functional }

We  begin by postulating general free energy functional broken into chemical and {\it nonlocal} electrostatic contributions. 
Let $G = G_{el} + G_{chem}$, where $G_{el}$ is the electrostatic energy and $G_{chem} = \int_V d\rb g$ is the chemical (non-electrostatic) part of the total free energy, $G$.  Suppose that $G_{chem}$ is known, and let us focus on electrostatic correlation effects in $G_{el}$.

The electrostatic potential,  $\phi$, is defined as the electrostatic energy per ion (free charge).  The  electrostatic energy cost for adding a charge $\delta \rho$ in the bulk liquid volume $V$ or $\delta q_s$ on the metal surface $S$ is,
\begin{equation}
\delta G_{el} = \int_V d\rb \, \phi \, \delta \rho + \int_S d\rb \, \phi \, \delta q_s.   \label{eq:dG}
\end{equation}
The charge is related to the displacement field $\Db$ via Maxwell's equation,
\begin{equation}
\nabla \cdot \Db = \rho \ \ \Rightarrow \ \ \delta \rho = \nabla \cdot \delta \Db.    \label{eq:max1}
\end{equation}
The corresponding boundary condition for an ideal metal surface (where $\Db=0$) is,
\begin{equation}
[\hat{n} \cdot \Db] = \hat{n}\cdot \Db = - q_s \ \ \Rightarrow \ \ \delta q_s = - \hat{n} \cdot \Db.   \label{eq:max2}
\end{equation}
Substituting these expressions into (\ref{eq:dG}) and using Gauss' theorem, along with the definition of the electric field, $\Eb = - \nabla\phi$, we recover the standard electrostatic free energy equation~\cite{landau},
\begin{equation}
\delta G_{el} = \int_V d\rb \, \Eb \cdot \delta \Db.  \label{eq:dG2}
\end{equation}

In the linear response regime (for small external electric fields), we have
\begin{equation}
\Db = \hat{\varepsilon} \Eb,
\end{equation}
where $\hat{\varepsilon}$ is a linear operator, whose Fourier transform $\hat{\varepsilon}(k)$ encodes how the permittivity depends on the wavelength $2\pi/k$ of the $k$-Fourier component of the field, due to discrete ion-ion correlations, as well as any non-local dielectric response of the solvent. A crucial feature of our approach, however, is that we do not restrict ourselves to small amplitude perturbations in Fourier space. Instead, we consider a general linear permittivity operator in real space and focus on correlation effects.  

By linearity, we can integrate  (\ref{eq:dG2}) over $\delta \Db$ through a charging process that creates all the charges in the bulk and surface from zero to obtain
\begin{equation}
G_{el} = \frac{1}{2} \int_V d\rb \, \Eb \cdot \Db.
\end{equation}
For a given distribution of charges $\rho$ and $q_s$, with associated displacement field $\Db$, the physical electric field $\Eb$ is the one that minimizes $G_{el}$, subject to the constraint of satisfying Maxwell's equations (\ref{eq:max1})-(\ref{eq:max2}). Since $\Eb = -\nabla\phi$ to enforce $\nabla\times\Eb=0$, we can minimize $G_{el}$ with respect to variations in $\phi$, using Lagrange multipliers $\lambda_1$ and $\lambda_2$ to enforce the constraints,
\begin{eqnarray}
G_{el}[\phi] &=& \int_V  d\rb \, \left[ \frac{1}{2} \Eb\cdot\Db+ \lambda_1 \left(\rho - \nabla\cdot \Db\right)\right] \nonumber \\
& & + \oint_S d\rb_s \, \lambda_2 \left( q_s + \hat{n}\cdot\Db \right).
\end{eqnarray}
To calculate the extremum, we use the
Fr\'echet functional derivative:
\begin{equation}
\frac{\delta G_{el}}{\delta \phi} = \lim_{\epsilon\to 0} \frac{ G_{el}[\phi + \epsilon \phi_0 \delta_\epsilon] - G_{el}[\phi] }{\epsilon \phi_0}
\end{equation}
where $\delta\phi_\epsilon = \phi_o \delta_\epsilon(\rb,\rb^\prime)$ is a localized perturbation of the potential (with compact support), which tends either to a 3D delta function in the liquid ($\rb \in V$) or to a 2D delta function on the surface ($\rb \in S$) as $\epsilon\to 0$, and $\phi_0$ is an arbitrary potential scale for dimensional consistency. By setting  ${\delta G_{el}}/{\delta \phi} = 0$ for both surface and bulk variations, we find $\lambda_1=\lambda_2=\phi$. Finally, using vector identities, we arrive at a general functional for the electrostatic energy,
\begin{equation}
G_{el}[\phi] = \int_V  d\rb \, \left( \rho\phi +  \frac{1}{2} \nabla\phi \cdot\Db \right) + \oint_S d\rb_s \, q_s \phi
\end{equation}
whose variational derivative with respect to $\phi$ will be set to zero, once we know the relationship between $\Db$ and $\Eb = -\nabla\phi$.

\subsection{ Nonlocal electrostatics for correlations }

To model the field energy, we assume linear dielectric response of the individual molecules (ions and solvent) with constant mean permittivity $\varepsilon$,  plus a simple non-local contribution for Coulomb correlations. Here, the permittivity $\varepsilon$ describes the electronic polarizability of the ions (for RTIL) as well as (in the case of electrolytes) the dielectric relaxation of the solvent.   There is an extensive literature on nonlocal electrostatic models of the form, $\Db(\rb) = \int d\rb' \varepsilon(\rb,\rb') \Eb(\rb')$, mainly focused on describing the nanoscale dielectric response of water~\cite{kornyshev1978,kornyshev1982,hildebrandt2004,rottler2009}. 
In this work, we take a very different approach because our aim is to model the transient formation of correlated ion pairs of opposite sign (zwitterions), which act as dipoles and contribute to the nanoscale dielectric response of strongly correlated ionic liquids.  

The theory begins by postulating a non-traditional form of the energy density stored in the electric field in the dielectric medium,
\begin{equation}
g_{field} = -\frac{1}{2} \nabla\phi \cdot\Db =  \frac{\varepsilon}{2}  \left( \Eb(\rb)^2 + \int_V d\rb^\prime K(\rb,\rb^\prime) \bar{\rho}(\rb)\bar{\rho}(\rb^\prime) \right)
\end{equation}
where we define 
\begin{equation}
\bar{\rho} = \varepsilon \nabla\cdot \Eb = -\varepsilon \nabla^2\phi,
\end{equation}
as the ``mean-field charge density'', which would produce same the electric field in the dielectric medium without accounting for ion-ion correlations~\cite{bazant2011}. In this theory, nonlocal electrostatic effects are assumed to derive from pairwise interactions between effective charges, defined in terms of the local divergence of the electric field via the standard second-order Poisson equation with constant permittivity $\varepsilon$. The non-local kernel $K(\rb,\rb^\prime)$ is intended to describe correlations between discrete pairs of fluctuating ions resulting from Coulomb interactions in the liquid.   

To take into account screening phenomena, we assume that the correlation kernel $K(\rb,\rb^\prime)$ decays exponentially over a characteristic length scale $\ell_c$. Below this distance, ions experience bare Coulomb interactions, and beyond it, thermal agitation and many-body interactions act to suppress direct electrostatic correlations.  The correlation length is clearly bounded below by the ion size $a$, which becomes the most relevant length scale in a highly concentrated electrolyte (including the solvent shell in the ion size) or a solvent-free ionic liquid. In the simplest version of our theory for dense ionic mixtures, it is possible to avoid adding any new parameter by simply setting $\ell_c=a$, as in our work on RTIL~\cite{bazant2011}. In dilute electrolytes, however, the correlation length should increase with the mean ion spacing, and we expect it to be cut off at the scale of the Bjerrum length $\ell_B$, which is the separation distance between ions below which the bare Coulomb energy exceeds the thermal energy $k_BT$.

In order to obtain a simple continuum model, we further assume that charge variations mainly occur over length scales larger than $\ell_c$ (corresponding to small perturbation wavenumbers, $\ell_c |k| \ll 1$). In this limit, we perform a  gradient expansion for the non-local term
\begin{equation}
g_{field} \sim \frac{\varepsilon}{2} \left[ |\nabla \phi |^2 +
\sum_{n=0}^\infty \alpha_{n}  \left( \frac{\ell_c^{n-1}}{\varepsilon} \nabla^n \bar{\rho}\right)^2  \right]
\end{equation}
where $\alpha_n$ are dimensionless coefficients, which implies
\begin{eqnarray}
G_{el}[\phi] &\sim& \int_V  d\rb \, \left\{ \rho\phi - \frac{\varepsilon}{2} \left[  |\nabla\phi|^2
+  \sum_{n=2}^\infty \alpha_{n-2} (\ell_c^{n-1} \nabla^{n} \phi)^2 \right] \right\} \nonumber \\
& & + \oint_S d\rb_s \, q_s \phi    \label{eq:Ggen}
\end{eqnarray}
For simplicity, we typically truncate the expansion after the first term, even though this may become inaccurate in situations of interest with charge density variations at the correlation length scale.

From the gradient expansion of the nonlocal electrostatic energy functional, we set ${\delta G_{el}}/{\delta \phi} = 0$ for bulk and surface perturbations in (\ref{eq:Ggen}).   In this way, we recover Maxwell's equations (\ref{eq:max1})-(\ref{eq:max2}), with 
\begin{equation}
\Db = \hat{\varepsilon} \Eb,
\end{equation}
where the permittivity operator has the following gradient expansion,
\begin{equation}
\hat{\varepsilon} = \varepsilon \left( 1-  \sum_{n=1}^\infty \alpha_{n-1} \ell_c^{2n} \nabla^{2n} \right).
\end{equation}
Eq. (\ref{eq:eps}) results from the first term in the gradient expansion  with the choice $\alpha_0=1$ (after suitably rescaling $\ell_c$), where the overall negative sign of this term is chosen to promote over-screening.
The corresponding small-$k$ expansion of the Fourier transform of the permittivity,
\begin{equation}
\hat{\varepsilon}(k) = \varepsilon \left[ 1 + \sum_{n=1}^\infty \alpha_{n-1} (-1)^{n-1} (\ell_c k)^{2n} \right] 
\end{equation}
grows with $k$ at small wavenumbers in the case where correlations promote overscreening, $\alpha_0>0$, as noted above. This is a hallmark of Coulomb correlations, promoting alternating charge ordering.

\section{Correlated electrokinetics at a planar surface}

\subsection{ Basic equations }
To demonstrate how correlation effects may influence double layer structure and
electrokinetic flows, we start by
exploring the behavior at a planar surface.  We assume
a 1D double layer at equilibrium with constant $\varepsilon$ and a  $z^+:z^-$ binary electrolyte.

The model we solve is thus,
\[
\varepsilon \left( \ell_c^2 \frac{d^4 \phi}{d x^4} - \frac{d^2 \phi}{d x^2}\right) = \rho = z^+ e c^+ - z^- e c^-.
\]
The boundary conditions at the electrode surface of fixed potential are,
\[
\phi = \phi_0,  \quad \frac{d^3 \phi}{d x^3} = 0.
\]
This electric potential equation is solved along with the equations
that the chemical potentials must be
constant at equilibrium.
In this work we consider the Bikerman model for volume constraints only with equal sized cations and anions
such that the chemical potential of the ions is,
\[
\mu_{\pm} = k T \mathrm{log} c_i - k T \mathrm{log}(1-a^3(c_+ + c_-) ) \pm z_\pm e \phi
\]

To calculate hydrodynamic slip, we start with the Navier Stokes equation and assume that in the electric
double layer we have a
balance between the electric body force and viscous forces,
\[
0=\eta \frac{d^2 u}{d x^2} + \rho_e E_t,
\]
where $E_t$ is the electric field tangential to the surface.
In our model this becomes,
\[
0=\eta \frac{d^2 u}{d x^2} +
\varepsilon \left( \ell_c^2 \frac{d^4 \phi}{d  x^4} - \frac{d^2 \phi}{d x^2} \right)E_t.
\]
As with the standard Helmholtz-Smoluchowski equation, we can integrate this
equation across the double layer twice to obtain (with the convention that far from the wall, $\phi$=0).
\[
u(\infty) =
-\frac{\varepsilon E_t \phi(0)}{\eta} \left( 1-  \left. \frac{\ell_c^2}{\phi(0)} \frac{d^2 \phi}{d x^2} \right |_{x=0}   \right).
\]
In the above expression, we are assuming that the medium permittivity and viscosity are constant within the
double layer, though this approximation can be relaxed.
An important general prediction is that the classical Helmholtz Smolukowski slip velocity, $U_{HS}  = -\varepsilon E_t \phi(0)/\eta$, is modified by the inclusion of correlation effects. This can be understood as a consequence of nonuniform permittivity.

The total charge in the double layer is given as the integral of the charge over the double layer,
\[
q = \int_0^\infty \rho_e dx  =
\int_0^\infty  \varepsilon \left( \ell_c^2 \frac{d^4 \phi}{d x^4} - \frac{d^2 \phi}{d x^2} \right) dx
\]
Evaluating this integral and using the boundary conditions at a solid electrode stated above
we obtain,
\[
q = \int_0^\infty \rho_e dx  =  \varepsilon \left. \frac{d \phi}{d x}\right|_{x=0},
\]
with the total capacitance defined as $C=q/\phi(0)$.

\subsection{Dimensionless formulation}
We assume a binary $z^+:z^-$ electrolyte such that the far field concentrations of the cations and anions
follows $z^+ c^+_\infty = z^- c^-_\infty$. For simplicity we assume that the cations and anions are of the same diameter.
We make the formulation dimensionless using the scales $\tilde{c}^+  = c^+/c^+_\infty$,
$\tilde{c}^-  = c^-/c^+_\infty$,  and  $\tilde{\phi} = \phi  (e/k T)$.
The dimensionless concentrations can be written as explicit functions of the electric potential,
\begin{equation}
\tilde{c}^+ = \beta(\tilde{\phi}) \mathrm{e}^{- z^+\tilde{\phi}}
\end{equation}
\begin{equation}
\tilde{c}^- =  \frac{z^+}{z^-}  \beta(\tilde{\phi}) \mathrm{e}^{ z^-\tilde{\phi}}
\end{equation}
where the function, $\beta$,  is given by
\begin{equation}
\beta(\tilde{\phi}) = \frac{1}{ 1 - \nu +  \frac{\nu}{z^- + z^+}\left(z^- \mathrm{e}^{-z^+\tilde{\phi}}
                                             + z^+ \mathrm{e}^{z^-\tilde{\phi}} \right)}
\end{equation}
 where $\nu$ is the volume fraction in the bulk and has a value $\nu=(1 + \frac{z^+}{z^-}) c^+_\infty a^3$.
 For the case of a 1:1 electrolyte note that $\beta(\tilde{\phi})=1/(1 + \nu(\mathrm{cosh}(\tilde{\phi}) -1))=1/(1 + \nu\mathrm{sinh}^2(\tilde{\phi}/2))$ as has been used in previous works \cite{kilic2007a,kornyshev2007}.
We relate the lattice size parameter, $a$, to the spherical ion diameter, $d$, as $a^3 = \frac{\pi}{6} d^3/0.63=0.83 d^3$ where the factor of 0.63 is the maximum volume fraction for random close packing of spheres.

The Poisson equation is
scaled by the Debye length; i.e. $\tilde{x} = x/\lambda_D$ where
\[
\lambda_D =\sqrt{\frac{\varepsilon k T}{e^2 c^+_\infty z^+(z^++z^-)}}.
\]
 Under this scaling our governing equation becomes,
\begin{equation}
\left(  \frac{d^2 \tilde{\phi}}{d \tilde{x}^2} -  \delta_c^2 \frac{d^4 \tilde{\phi}}{d \tilde{x}^4}\right) =
\beta(\tilde{\phi}) \frac{\mathrm{e}^{-z^+ \tilde{\phi}} - \mathrm{e}^{z^- \tilde{\phi}}}{ (z^+ + z^-)}
\end{equation}
where $\delta_c = \ell_c/\lambda_D$.
 This equation is subject to the boundary conditions that the potential at the electrode is fixed, the third derivative  of the potential is zero, and the potential goes to zero smoothly at infinity.

 There are three dimensionless parameters which emerge from  our formulation, the bulk volume fraction $\nu$, the
correlation length scale $\delta_c$, and the applied voltage (or known surface charge).
The solution also depends on the valences of the ions $z^+$ and $z^-$.

 In dimensionless terms, the slip velocity relative to the
 Helmholtz-Smulokowski velocity is,
\begin{equation}
    \frac{u(\infty)}{U_{HS}} = \left( 1 -   \left. \frac{\delta_c^2}{\tilde{\phi}(0)} \frac{d^2 \tilde{\phi}}{d \tilde{x}^2} \right |_{\tilde{x}=0} \right).
\end{equation}
where $U_{HS} = \varepsilon E_t \phi(0)/\eta$. The capacitance relative to the Debye-Huckel capacitance, $C_{DH}=\varepsilon/\lambda_D$ is simply,
\begin{equation}
    \frac{C}{C_{DH}} = \left. - \frac{1}{\tilde{\phi}(0)} \frac{d \tilde{\phi}}{d \tilde{x}}\right|_{\tilde{x}=0}.
\end{equation}
For the remainder of the paper we will drop the tilde notation and only refer to dimensionless quantities in our equations.

\subsection{Low voltage analytical solutions}

When the voltage is small relative to the thermal voltage, $k T/e$, the problem is
drastically simplified and the right hand side of our equation becomes,
\begin{equation}
  \left( \frac{d^2 \phi}{d x^2} - \delta_c^2 \frac{d^4 \phi}{d x^4} \right) =  \phi.
\end{equation}
This equation can be solved analytically, though the form depends upon whether the value of
$\delta_c$ is less than, greater than, or equal to $1/2$.

\subsubsection{Solution $\delta_c<1/2$, "weak" correlation effects}
When $\delta_c<1/2$ the analytical solution at low voltage has the form,
\begin{equation}
\phi(x) = \frac{\phi(0)}{1-k_1^3/k_2^3} \left( e^{-k_1 x}   - \frac{k_1^3}{k_2^3} ~ e^{-k_2 x} \right),
\end{equation}
where
\[
k_1 = \sqrt{ \frac{1 - \sqrt{1-4 \delta_c^2}}{2 \delta_c^2}}, ~~
k_2 = \sqrt{ \frac{1 + \sqrt{1-4 \delta_c^2}}{2 \delta_c^2}}.
\]
The capacitance of the double layer is,
\begin{equation}
\frac{C}{C_{DH}} =  \frac{ k_1 \left( 1 - \frac{k_1^2}{k_2^2}  \right) }{1-k_1^3/k_2^3},
\end{equation}
and the  slip velocity is,
\begin{equation}
\frac{u(\infty)}{U_{HS}} = \left(   1- \delta_c^2 \frac{ k_1^2 \left(1   - \frac{k_1}{k_2}  \right)}{1-k_1^3/k_2^3}   \right).
\end{equation}
In the limit of $\delta_c$ going to zero $k_1=1$ and $k_2 = \infty$, thus we
recover the  Debye-Huckel solution $\phi(x) = \phi(0) e^{-x}$.
This new solution has a functional form very similar to the classic double layer.
The structure is given as the sum of two exponentials
with decay lengths on the order of unity, though slightly modified.

\begin{figure*}
a) \hspace{2.3in}  b)\hspace{2.3in}  c)\\
 \includegraphics[width=2.3in]{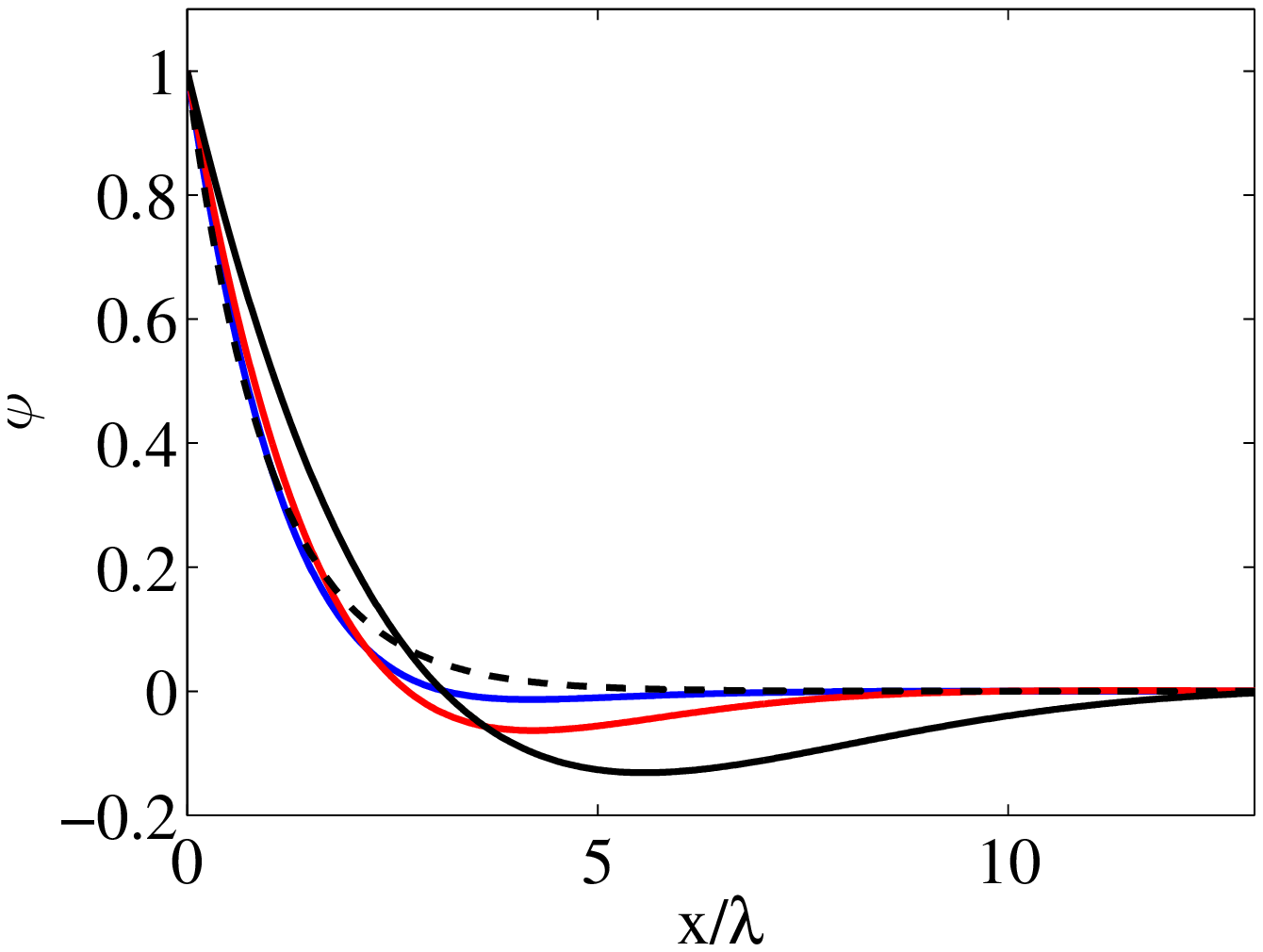}
 \includegraphics[width=2.3in]{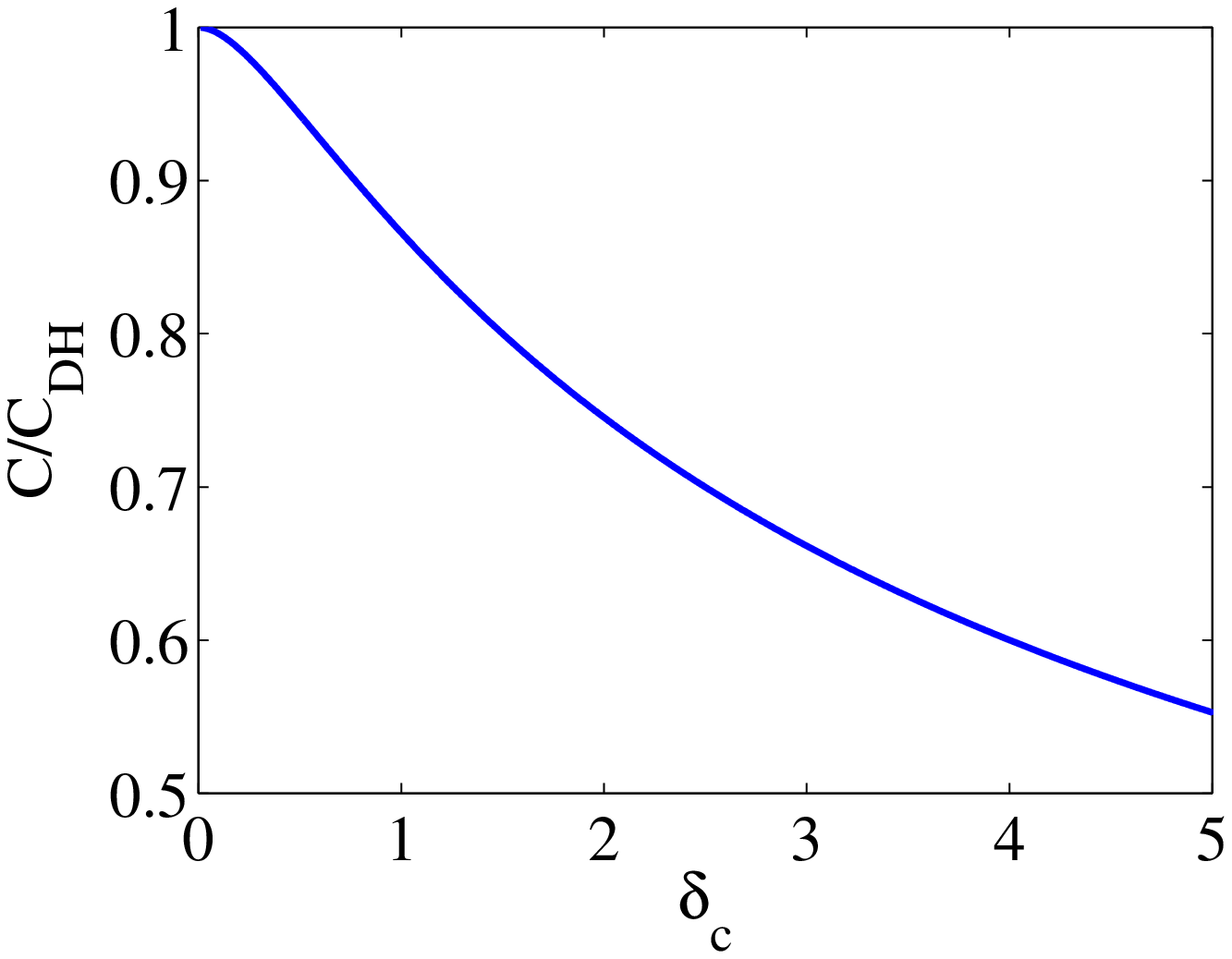}
 \includegraphics[width=2.3in]{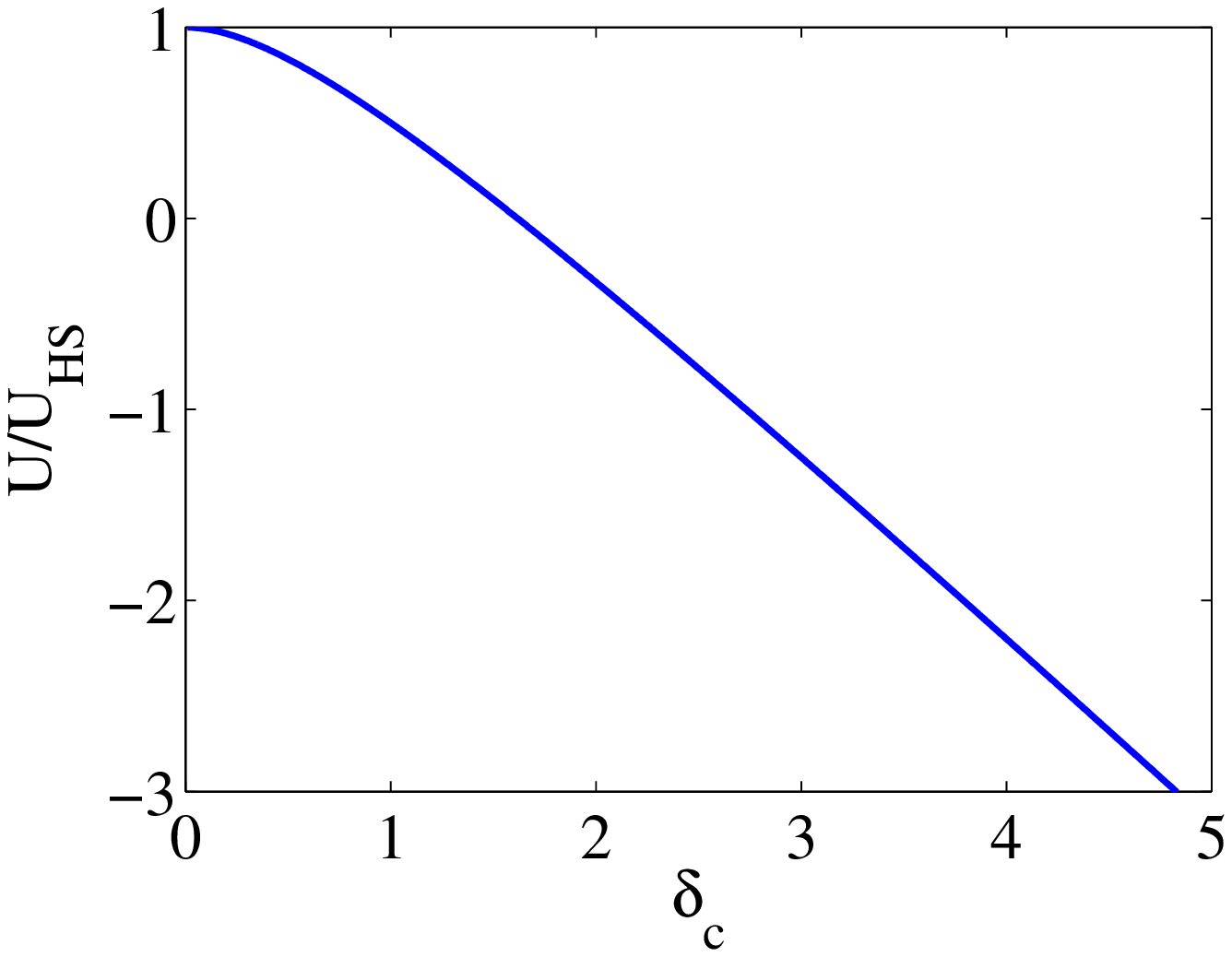}
\caption{Low voltage solutions to the continuum model. a) Double layer structure at different values of $\delta_c$. Solutions are shown for $\delta_c=0$ (dashed) and
$\delta_c=1,2,$ and $5$.
b) Capacitance and c) slip velocity  as a function of the
correlation length scale, $\delta_c$.
}
\label{fig:dh}
\end{figure*}

\subsubsection{Solution for $\delta_c>1/2$, "strong" correlation effects}
When $\delta_c>1/2$ the analytical solution at low voltage has the form,
\begin{equation}
\phi(x) = \phi(0)  e^{-k_1 x}  \left( \mathrm{cos}(k_2 x) - A ~\mathrm{sin}( k_2 x) \right)
\end{equation}
where
\[
k_1 = \frac{\sqrt{2 \delta_c + 1}}{2 \delta_c}, ~
k_2 = \frac{\sqrt{2 \delta_c - 1}}{2 \delta_c}, ~
A  = \frac{\sqrt{2 \delta_c + 1} (\delta_c -1)}{\sqrt{2 \delta_c - 1} (\delta_c + 1) }
\]
The capacitance of  the double layer is,
\begin{equation}
\frac{C}{C_{DH}} = \frac{\sqrt{2 \delta_c + 1}}{ \delta_c+1}
\end{equation}
which decays with increasing correlations. The 
slip velocity is,
\begin{equation}
\frac{u(\infty)}{U_{HS}} =
  \left(   1-    \frac{\delta_c^2}{\delta_c +1} \right).
\end{equation}
The slip velocity changes sign if $\delta_c$ is sufficiently large. In particular, there is an  {\it electro-osmotic  flow reversal} or  {\it electrokinetic charge inversion} of the surface when the dimensionless correlation length exceeds the golden mean:  $\delta_c > (1+\sqrt{5})/2 = 1.618$.

The form of the double layer becomes modified as $\delta_c$ increases. We find that the functional form consists of decaying
sinusoids with a length scale provided explicitly by $k_1$ and $k_2$.
At relatively large values of $\delta_c$ the length scale of the decay and the oscillations is approximately $\sqrt{2 \delta_c}$.

\subsection{ Numerical results   }
At low voltage, the solution has only one free parameter, the correlation length scale, $\delta_c$.
The structure of the double layer as $\delta_c$ is varied is shown in Figure \ref{fig:dh}. We see that as the strength of the
correlations is increased the double layer shows charge density oscillations.
From the analytical solution we see that the oscillations emerge when $\delta_c$ is greater than 1/2.
The length scale for the whole double layer also increases  as the correlations are increased.
 From the analytical solution we can easily see at large $\delta_c$ that the size of the double layer grows
 with the square root of $\delta_c$.
For small values of $\delta_c$  the results  become indistinguishable
from the classic Debye-Huckel solution.

In Figure \ref{fig:dh} (b)  we show the capacitance and (c) slip velocity as a function of $\delta_C$.
We see a decrease in the slip velocity and the capacitance with increasing $\delta_c$.
As  correlation effects  become stronger  the
flow is quenched and then reverses direction.
Note that from the analytical solution that at $\delta_c=1$
that the flow velocity is half of $U_{HS}$ and the flow reverses when  $\delta_c>0.618$.
These values of $\delta_c$ are easily reached at high concentration in aqueous electrolytes, as we will soon see.

\begin{figure*}
a) \hspace{2.3in}  b)\hspace{2.3in}  c)\\
 \includegraphics[width=2.3in]{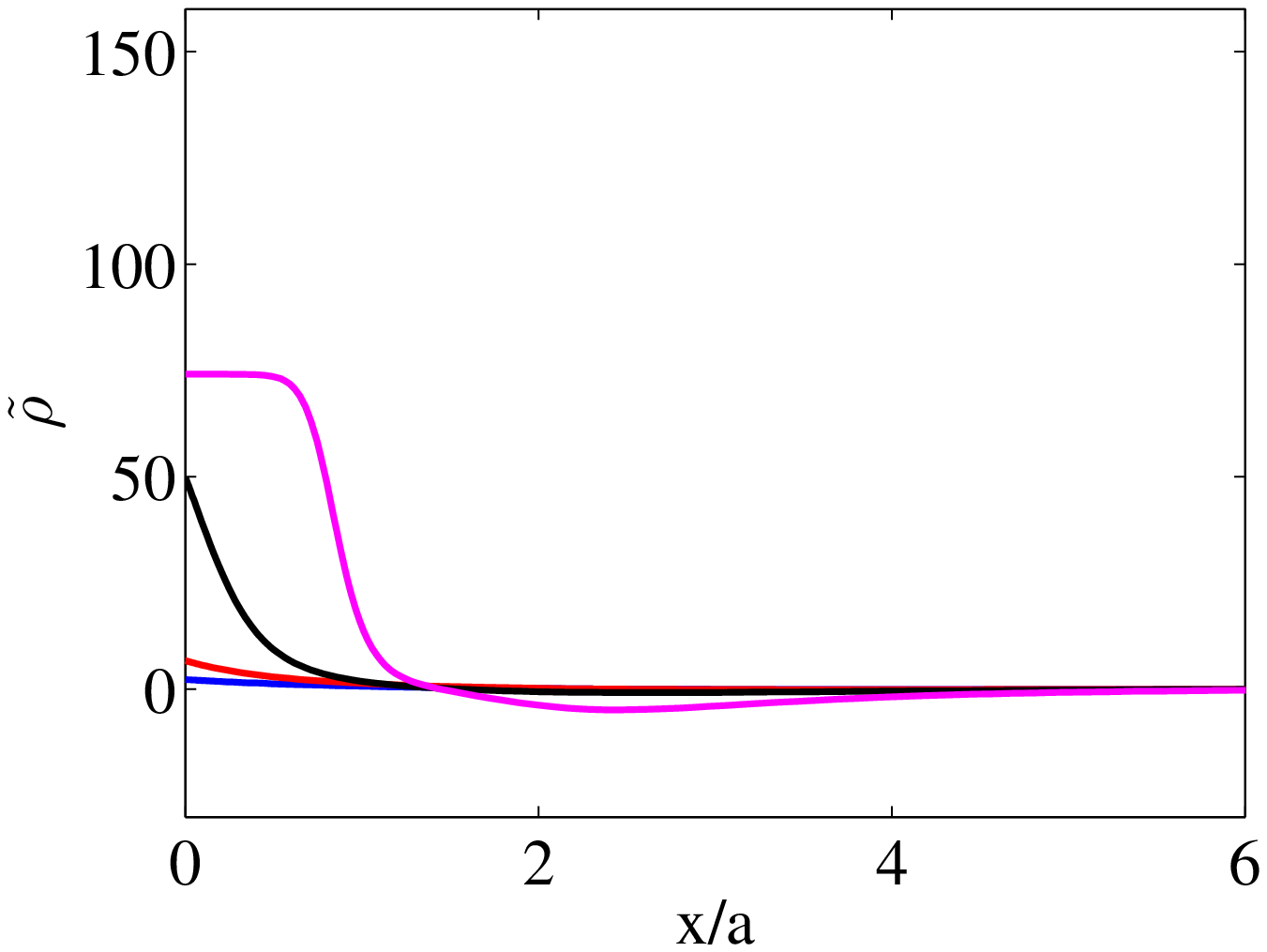}
 \includegraphics[width=2.3in]{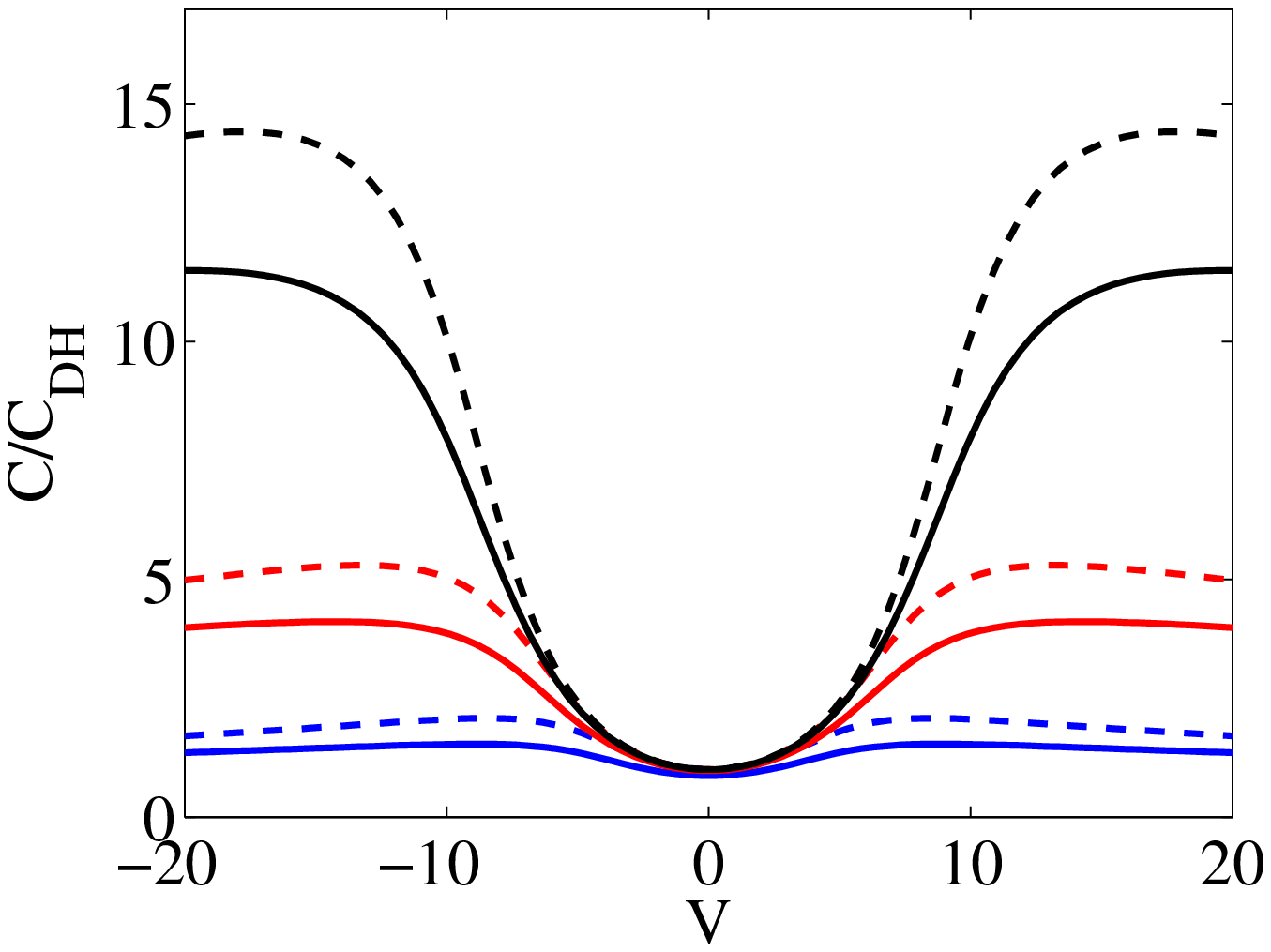}
 \includegraphics[width=2.3in]{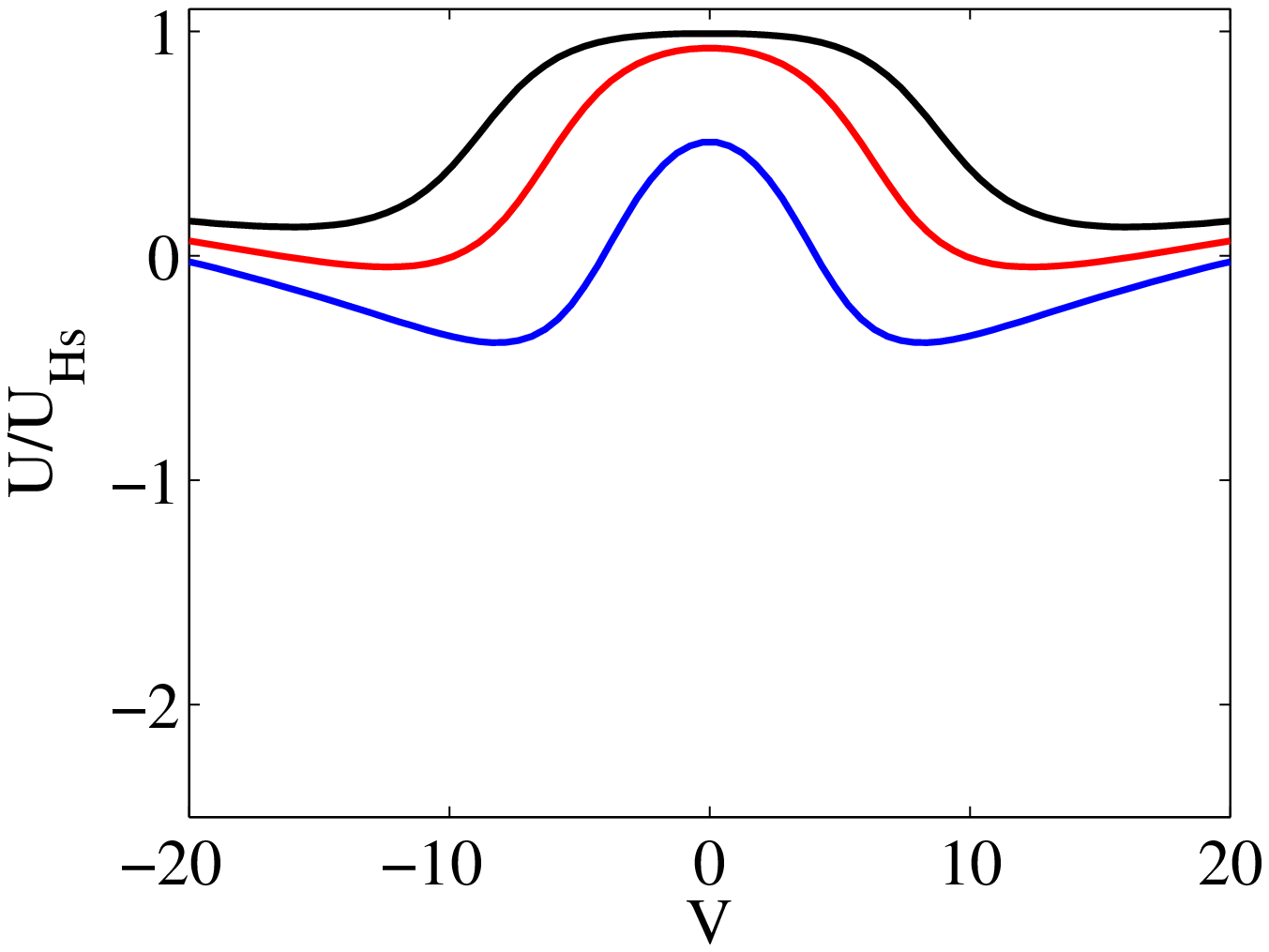}\\
 d) \hspace{2.3in}  e)\hspace{2.3in}  f)\\
 \includegraphics[width=2.3in]{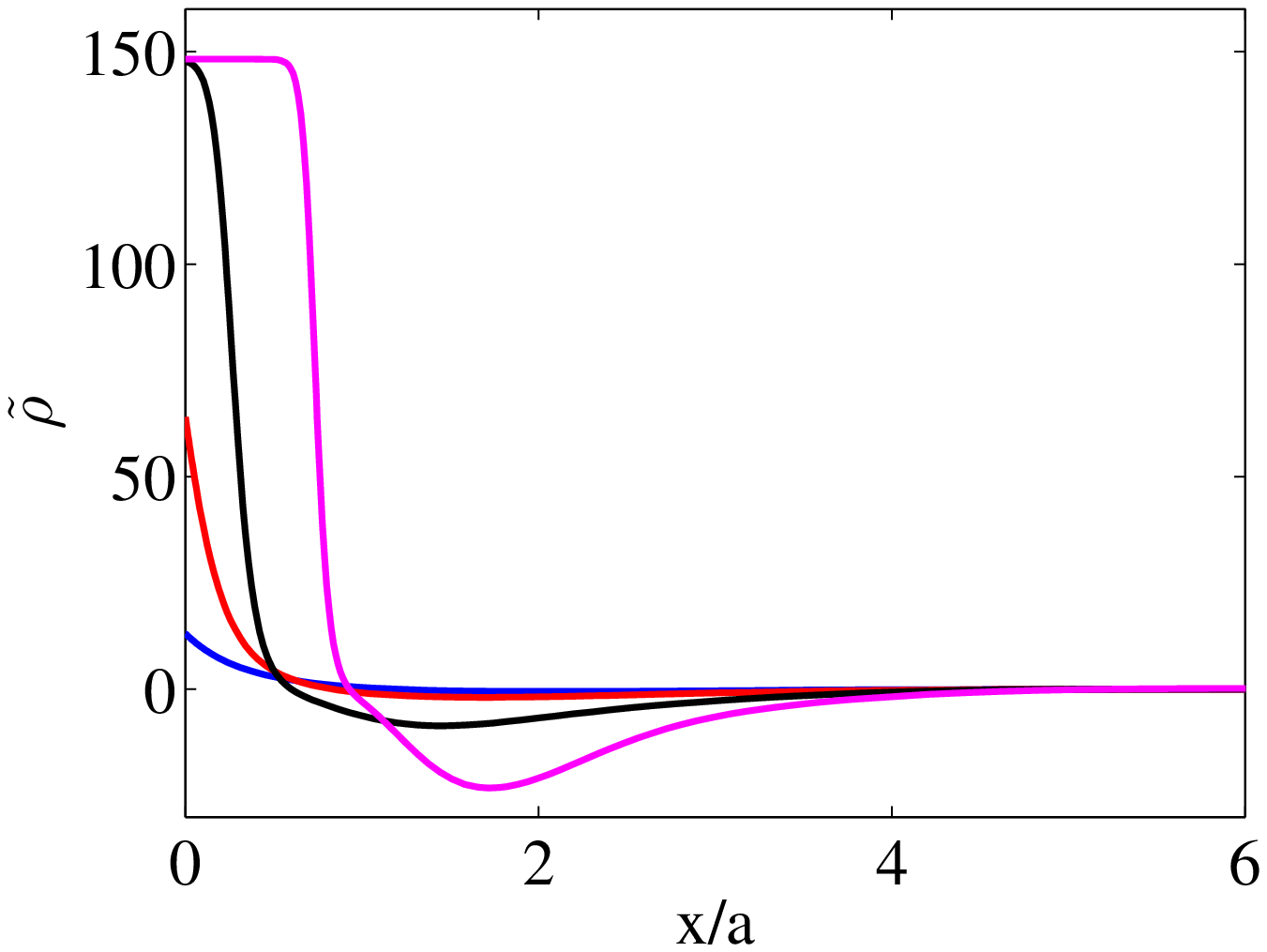}
 \includegraphics[width=2.3in]{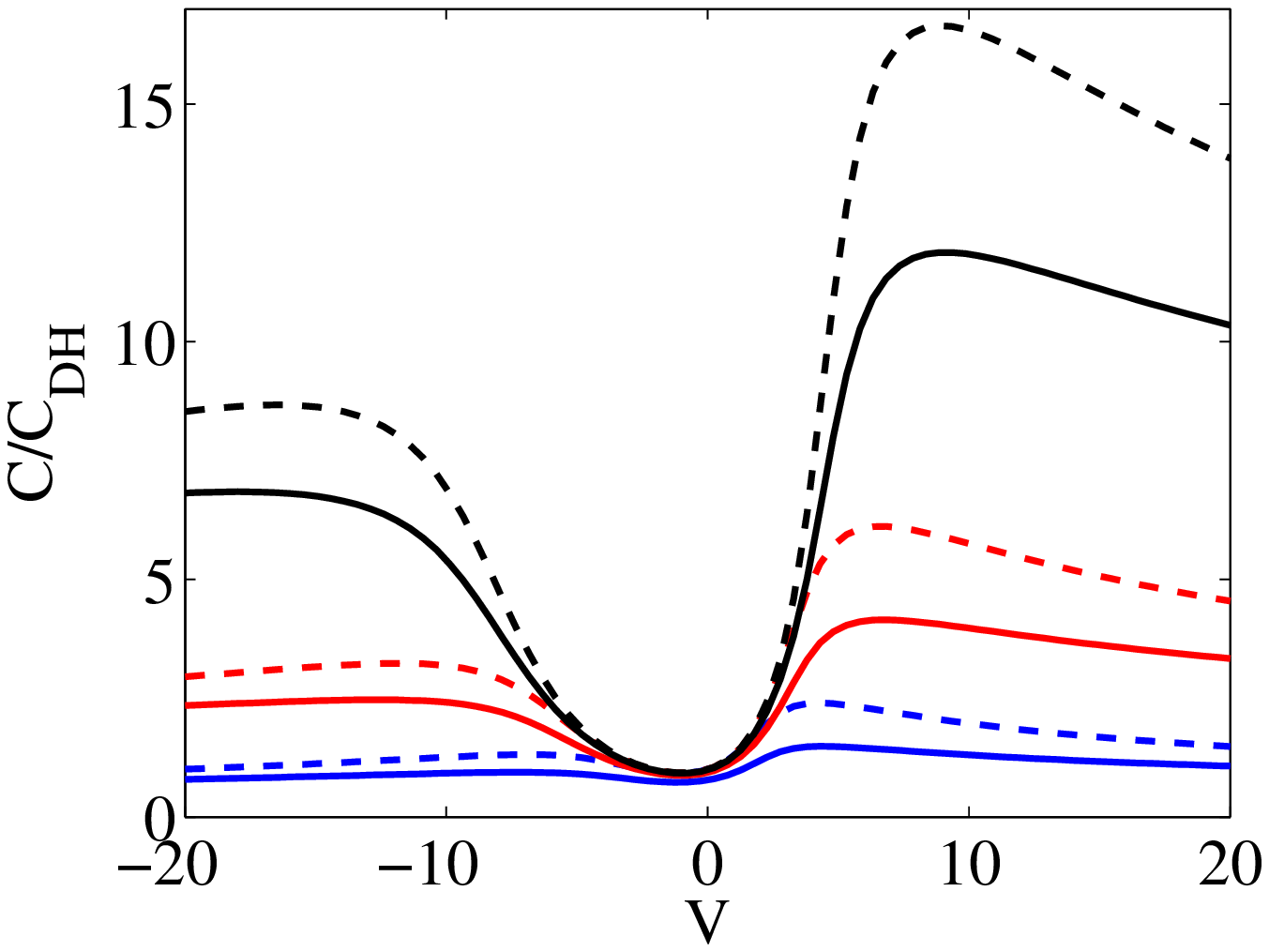}
 \includegraphics[width=2.3in]{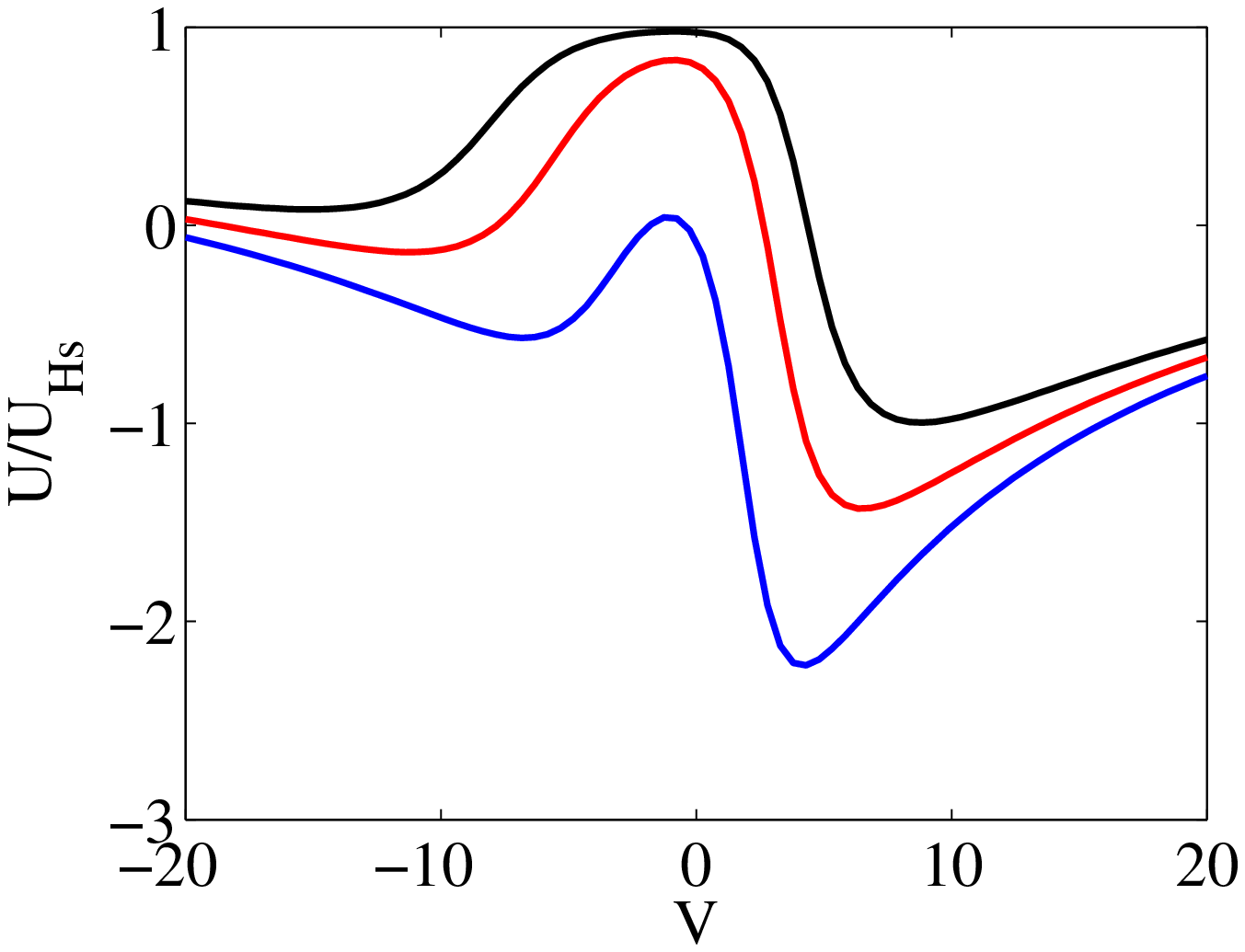}
\caption{Example solutions for a 1:1 (a, b, and c) and 2:1 electrolyte (d, e, and f) with 0.3 nm ion sizes. (a) Double layer structure showing the charge density profiles at wall voltages of -1, -2, -5 and -10 in units of $k_b T/e$ for a 1 Molar concentration of cations.
(b) Dimensionless capacitance as a function of voltage for concentration of 0.01 (black), 0.1 (red) and 1 (blue) molar from top to bottom. Corresponding solutions with no correlation effects ($\delta_c=0$) are shown as the dashed lines.  (c) Dimensionless slip velocity as a function of voltage for concentration of 0.01 (black), 0.1 (red) and 1 (blue) molar. Without correlations the slip velocity is always 1.
Figures (c), (d), and (e) are the same, only for the 2:1 system. The asymmetry is easily seen in the capacitance and slip velocity.
}
\label{fig:voltage}
\end{figure*}

At higher applied voltage the structure of the solution changes dramatically as we show in Figure \ref{fig:voltage}.
Here we show sample solutions for a 1:1 and  2:1 electrolyte of 0.3 nm ions as the voltage is changed.
In Figure \ref{fig:voltage}a we show the structure of the double layer at different voltages at a cation concentration of 1 molar. Using the ion size as the correlation length scale and as the volume fraction then
for the 1:1 system
$\delta_c=0.988$, $\nu=0.0270$
and for the 2:1 system
$\delta_c=1.71$,  $\nu=0.0405$.
 As the voltage increases, the charge density at the wall saturates to a value determined by the steric constraints. This condensed layer of ions grows as the voltage is increase. Without the correlations effect the charge density would decay monotonically from the maximum value to zero far from the wall. However, with the correlation effects included in the model, the charge density oscillates and changes sign.
 These oscillations are more pronounced in 2:1 system when the divalent ions crowd the wall.

 Turning to the capacitance in Figure \ref{fig:voltage}b we find that correlation effects reduce the  capacitance.
 The dimensionless capacitance is always 1 at zero voltage when $\delta_c=0$, however when  $\delta_c>0$ the  capacitance at zero voltage reduces according to Figure \ref{fig:dh}b.
 At higher voltage, the shape of the capacitance curve is similar to when $\delta_c=0$, the values are simply lower.
 This reduction in capacitance is consistent with previous work on steric constraints with the Bikerman model which found generally that the theory needed ion sizes that were bigger than one would expect physically to fit the experimental data \cite{storey2008,large_acis}

 The most dramatic departure from the classical model comes when computing the slip velocity in Figure \ref{fig:voltage}c. We see that at high concentration the model can predict reverse flow even at small voltages in the 2:1 system.
 At low concentration, we find that the model predicts classic slip  at low voltage but predicts reverse flow as the voltage is increased even  moderately. As the voltage is increased further, the model predicts the forward component of the flow begins to increase as the condensed layer grows. At high voltage the slip velocity for all concentrations begin to come together as the condensed layer begins to dominate the double layer structure.

 These preliminary flow results must be interpreted with caution. The model currently does not account for changes in the viscosity of the solution near the wall in the condensed layer. It is also unclear (as it is in classical theory) where the slip plane should be placed.  
 Recent work by Jiang and Qiao shows via molecular dynamics simulations that electroosmotic flow can be amplified by short wavelength hydrodynamics \cite{Jiang2012}. These effects (and others) are not included in our model and may be required for more detailed comparisons with experimental data.  

\section{ Validation}

\subsection{ Comparison with molecular simulations }

In order to determine whether this model has validity in the context of aqueous electrolytes, we can
compare the model predictions to those made by more sophisticated simulations such as Monte Carlo or Density Functional Theory (DFT). Monte Carlo simulations are often considered the standard  for equilibrium
chemistry while DFT has proven to quantitatively compare well against Monte Carlo at a much lower computational cost \cite{Gillespie2011}. Our aim is to determine whether an even  simpler continuum model can capture the same features.

In a prior paper we compared this correlations model to molecular dynamics simulations of ionic liquids \cite{bazant2011}.
In that work we assumed that the potential at $x=0$ in the continuum theory was the potential offset from the wall
by the radius of the ion. In comparisons to data for electrolyte solutions that follow, we find that
here we obtain good results by taking the voltage at $x=0$ to be the electrode, i.e. not accounting for the radius
of the ion as it approaches the surface.

\begin{figure}
\includegraphics[width=3.3in]{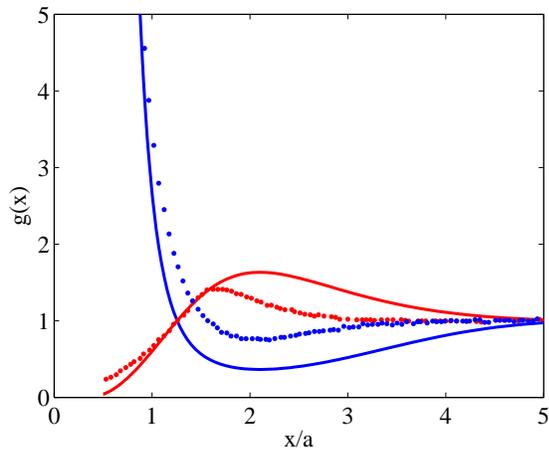}
\caption{Comparison of the continuum model (solid lines) to Monte Carlo simulations of ref \cite{Boda2002}.
The conditions here are a 2:1 electrolyte with surface charge of -0.3 $\mathrm{C/m^2}$ and an ion diameter of 0.3 nm. The points are the Monte Carlo simulation and the solid lines are the continuum model.
}
\label{fig:dist}
\end{figure}

In Figure \ref{fig:dist} we compare the ion distributions, $g(x) = c(x)/c_\infty$, predicted by the continuum
model to those predicted by Monte Carlo simulations of Boda et al \cite{Boda2002}. The conditions here are a 2:1 electrolyte with surface charge of -0.3 $\mathrm{C/m^2}$ and an ion diameter of 0.3 nm.
We find that the continuum model predicts much of the same structure as the Monte Carlo
simulations, though the length scale of the oscillations and the amount of over-screening predicted by the continuum model is larger than seen in the simulations. Better agreement can be obtained by reducing the correlation length scale by about 50 percent.  However, the classic electrokinetic model can only predict ion profiles which decay monotonically, so it is interesting that this extension for correlations effects  can provide the basic double layer structure with no  fitting parameters.

\begin{figure}
a) \includegraphics[width=3.3in]{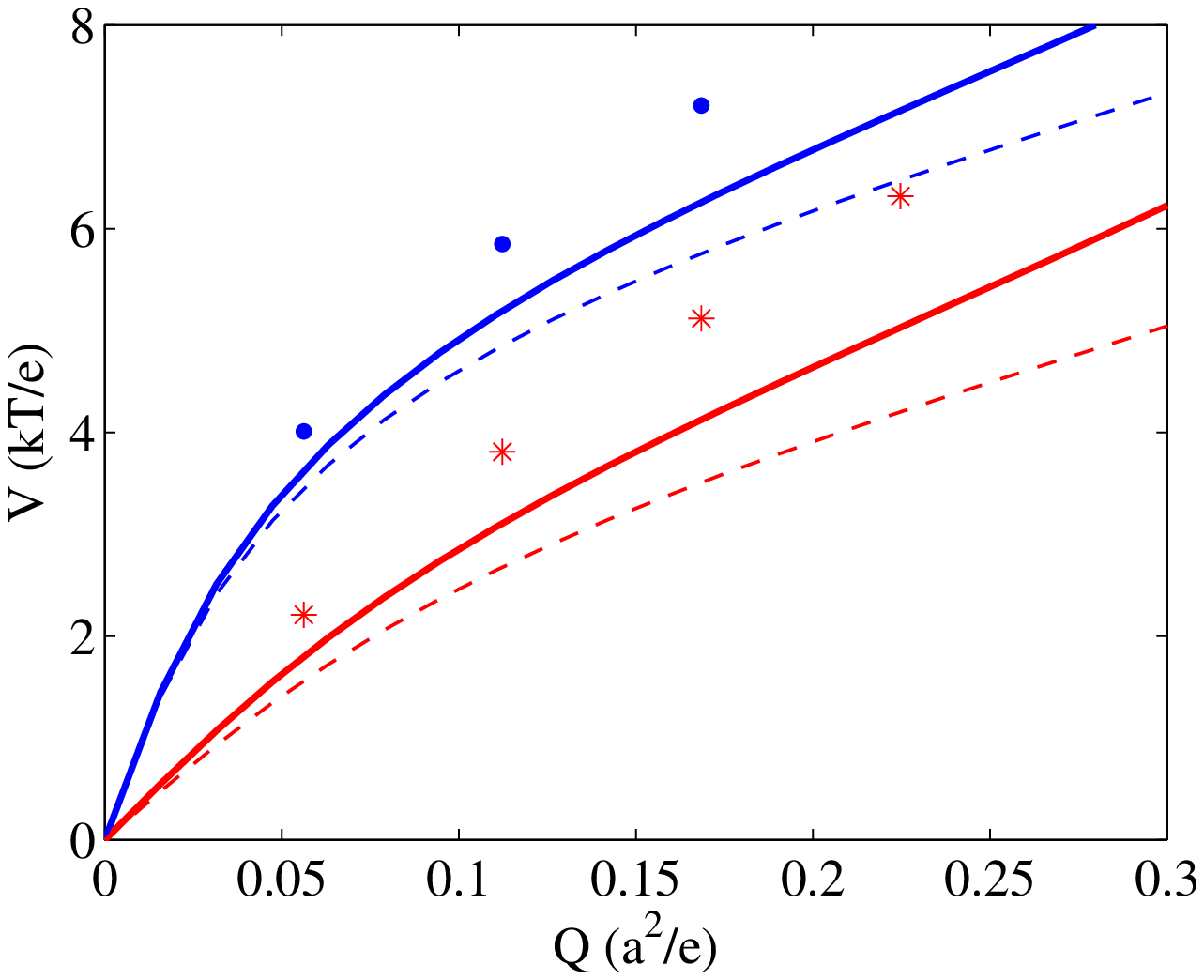}
b) \includegraphics[width=3.3in]{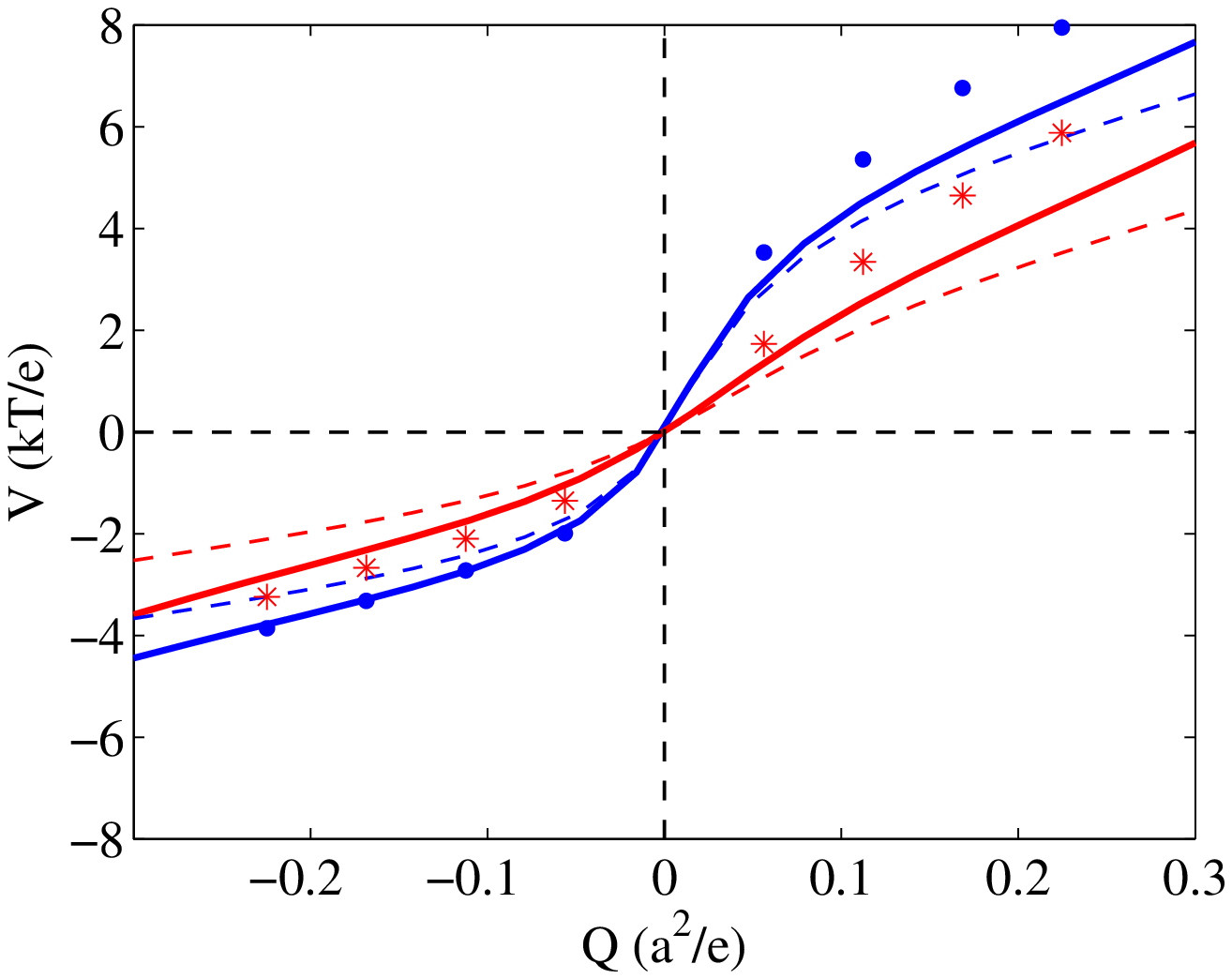}
\caption{Comparison of the continuum model with correlations (solid lines) to Monte Carlo simulations of ref \cite{Boda2002} (points) and the continuum model with only steric effects (dashed lines).
The ion diameter is 0.3 nm. In a) we show the result for a 1:1 electrolyte and in b) we show the result for a
2:1 electrolyte.
The upper solid blue curve and dots is for 0.1 M and the lower red curve and '*' is for 1 M concentration.
}
\label{fig:qv}
\end{figure}

In Figure \ref{fig:qv} a-b we compare the continuum model to the Monte Carlo simulations looking at the relationship between  the double layer charge and electrode  voltage. In a) we show results for a monovalent ion and in b) we show a 2:1 electrolyte at two different concentrations. The continuum model predicts the basic trends of the more complicated MC simulations, though under-predicts the voltage for a given charge.
The inclusion of correlation effects brings the continuum results  in better agreement with the MC simulations than when we only account for finite size effects.

\begin{figure}
 \includegraphics[width=3.3in]{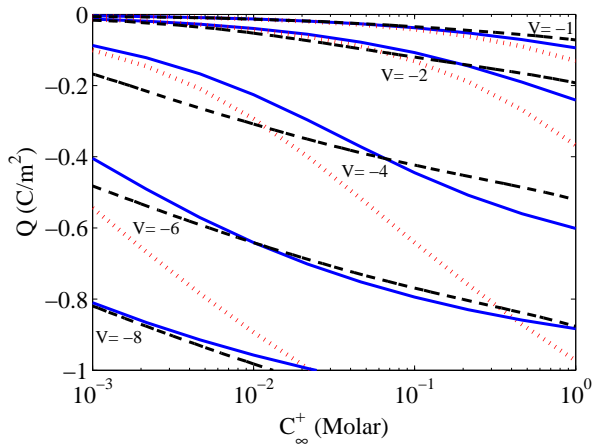}
\caption{Comparison of the continuum model accounting for correlation effects (blue solid lines)
to the DFT simulations of Gillespie et al \cite{Gillespie2011} (black dashed)
to the continuum model with $\delta_c=0$ (red dotted lines).
The ion diameter is 0.3 nm in the models and DFT.
}
\label{fig:dft}
\end{figure}

In Figures \ref{fig:dft} we compare the continuum model to results of density functional theory (DFT) simulations of Gillespie et al \cite{Gillespie2011} for a 2:1 electrolyte. In  Figures \ref{fig:dft} we show curves of constant voltage over a range of surface charge and concentration.
The results with the continuum model are in reasonable agreement with the DFT results, especially at large concentrations and high charge.
Importantly, the shape of these curves computed with DFT
are well predicted by this simple continuum model.  When $\delta_c=0$ and at high concentration, the continuum  model qualitatively departs  from the DFT results.
What is interesting about the continuum model with correlations included is that there are no fit parameters.

\subsection{ Comparison with experiments }

\begin{figure}
\includegraphics[width=3.3in]{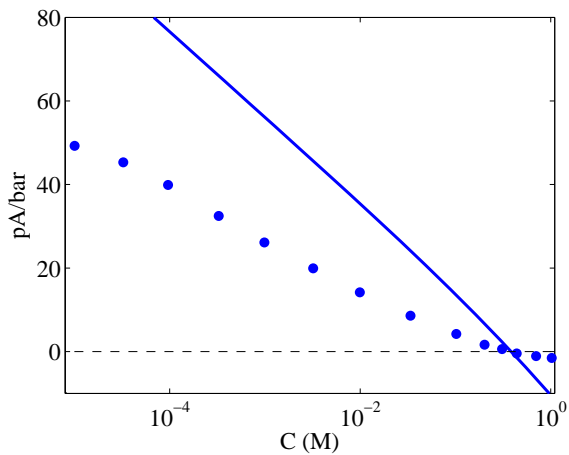}
\caption{Comparison of the continuum model experimental nanochannel data of  \cite{vanderH2006}. The electrolyte is a 2:1 with an assumed ion size of 0.3 nm.
}
\label{fig:exp}
\end{figure}

We can also compare the model to an experiment, rather than to other simulations, as a more definitive test. In Figure \ref{fig:exp} we compare the model to the nanochannel electrokinetic data of van der Heyden et al.  \cite{vanderH2006} as was done by
 Gillespie et al \cite{Gillespie2011}. In the experiment a nanochannel with a characterized surface charge is  driven by a pressure driven flow and the streaming current is measured. In this case the flow is driven by pressure and not electro-osmotically. To compute the streaming current we simply multiply our charge density profiles by the pressure driven velocity profile;
\begin{equation}
I = W \int_0^H \rho(x) u(x) dx
\end{equation}
where $W$ is the channel width of 50 $\mu$m, $H$ is the channel height of 450 nm, and
$u$ is the parabolic velocity profile. Since the double layer is so thin  relative to the channel height of 450 nm, we can safely assume that the pressure driven velocity profile is locally linear at the wall;
$du/dx=4 \Delta P H/(L \eta)$ for Pouiselle flow. Thus to compute the current per unit pressure drop for pressure driven flow we calculate,
\begin{equation}
\frac{I}{\Delta P} = \frac{4 W  H   }{L \eta} \int_0^\infty \rho(x) x dx.
\end{equation}
The current  per unit pressure  as a function of concentration is plotted in Figure \ref{fig:exp}
comparing the continuum model to the experiment. The agreement is qualitatively correct and predicts a reversal in the current around the same concentration as seen in the experiments.
The slower velocities at high concentration seen in the experiment is consistent with charge induced thickening, and increase in viscosity in a condensed layer of ions \cite{large_acis}.
There is still uncertainty in application of this model for flow. It is unclear where the slip plane should sit and whether the solution viscosity near the wall should be taken as a constant.
This uncertainty applies equally to the continuum model and the DFT simulations, as in those simulations the
current is calculated in the same way it is here, only the charge profile is calculated via DFT in their work is used.
 More experimental data under controlled situations is needed for further testing predictions of flow.

We also briefly draw attention to induced-charge electro-osmotic flows (ICEO)~\cite{large_acis,cocis2010}, where the new model may help to explain some puzzling experimental results (although we do not report any new simulations here). In particular, we (along with L. R. Edwards and M. S. Kilic) showed that flow reversal in AC electro-osmotic micro-pumps (consisting of  interdigitated planar micro-electrode arrays) could be explained by a Bikerman-like model of the double layer, where the differential capacitance of the double layer decreases at high voltage~\cite{storey2008}. A difficulty with this interperation of the experimental data, however, was the fact that the inferred ion size was far too large, whether considering a lattice gas or hard spheres. The problem could be alleviated by considering the possibility of reduction of the dielectric constant near the surface, and we speculated that correlation effects might further reduce the effective ion size in the model.  From the present work, we can see that electrostatic correlations tend to reduce electro-osmotic flow while also lowering the double-layer differential capacitance. The former effect could be wholly or partly misinterpreted as a sign of charge-induced thickening (i.e. an increase in viscosity in a highly charged double layer that could also reduce the net electro-osmotic flow), while the latter could reduce the capacitance without invoking such strong steric effects with large effective ion sizes.  Based on this evidence, it seems plausible that the new model might help to describe ICEO flows at high voltage and high salt concentration, which have otherwise resisted a complete theoretical understanding~\cite{large_acis}. 

\section{Conclusions}
We have developed a continuum model for electrokinetic phenomena that accounts for electrostatic correlations and applied this model to electro-osmotic flow and streaming current in aqueous electrolytes of high valence and high salt concentration at a flat, homogeneously charged surface.
The model predicts the basic electric double layer structure that has been observed in
Monte Carlo simulations; namely oscillations in the charge density and reversal of apparent charge of a surface based on electrokinetic flow.
Without any fitting parameters, the continuum model which also includes finite ion size effects reproduces features of much more complicated theories and simulations.  While the quantitative agreement between this model and Monte Carlo or DFT simulations is only approximate, the trends are much closer than found with the classical mean-field theory.
As in the case of RTIL~\cite{bazant2011}, it is remarkable that such a simple continuum theory can predict various subtle aspects of double layer structure and electrokinetic phenomena at the molecular scale.

\section*{Acknowledgments}
This work was supported by the National Science
Foundation, under Contracts No. DMS-0707641
(M. Z. B.) and No. CBET-0930484 (B. D. S.). We thank D. Gillespie for providing the raw DFT data used in
to compare our model against in Figure \ref{fig:dft} and for helpful discussions.

\bibliography{elec28}


\end{document}